\documentclass[a4paper,twocolumn,aps,prl]{revtex4}

\newcommand{\eq}[1]{\begin{equation}#1\end{equation}}
\newcommand{\dd}{\mathrm{d}}
\newcommand{\ee}{\mathrm{e}}

\newcommand{\Ai}{\mathrm{Ai}}
\newcommand{\Tr}{\mathrm{Tr \,}}
\usepackage{amsmath}
\usepackage{graphics}
\usepackage{graphicx}
\usepackage{psfrag}

\begin{document}

\title{Full counting statistics in a propagating quantum front and random matrix spectra}

\author{Viktor Eisler}
\affiliation{Vienna Center for Quantum Science and Technology,
Faculty of Physics, University of Vienna,
Boltzmanngasse 5, A-1090 Wien, Austria}
\author{Zolt\'an R\'acz}
\affiliation{Institute for Theoretical Physics - HAS,
E\"otv\"os University, P\'azm\'any s\'et\'any 1/a, 1117 Budapest, Hungary}

\date{\today}

\begin{abstract}

One-dimensional free fermions are studied with emphasis on propagating fronts emerging from a step initial condition.
The probability distribution of the number of particles at the edge of the front is determined exactly. It is found that
the full counting statistics coincides with the eigenvalue statistics of the edge spectrum of matrices from the Gaussian
unitary ensemble. The correspondence established between the random matrix eigenvalues and the particle positions
yields the order statistics of the right-most particles in the front and, furthermore, it implies their subdiffusive spreading.

\end{abstract}

\maketitle

The theory of quantum noise progressed rapidly during the last decades and
it became an important research area in the studies of transport in mesoscopic systems \cite{BBrev,Nazarov}.
One of the key concepts of this field is the full counting statistics (FCS), giving the probability
distribution of the charge transmitted through a conductor \cite{LL93}.
Non-Gaussian fluctuations of the FCS carry important information and the measurement of
higher order cumulants has recently become accessible in a number of experiments
on tunnel junctions \cite{RSP03,Bomze05}, quantum dots \cite{FHTH06,Gustavsson06}
and quantum point contacts \cite{GBSR08}.
The FCS has also been proposed as a tool to extract the entanglement entropy \cite{CCD09}
in terms of the measured cumulants \cite{KL09,Song11,CMV12}.

Theoretically, the FCS is best understood for the transport of non-interacting electrons between
conductance channels separated by a scatterer \cite{LL93}.
At zero temperature, an even simpler setup is to consider a one-dimensional system
with perfect transmission, where the current is induced  by preparing an initial state with
two parts of the system having unequal particle densities. 
In this case, the complete time evolution of the FCS can be followed \cite{Klich03, Schoenhammer07}
and interesting results (e.g. particle number fluctuations logarithmically growing with time
\cite{Schoenhammer07,AKR08}) can be obtained. 

The above setup with a step-like initial density is also remarkable because, due to the
density bias, a front builds up and penetrates ballistically into the low density region \cite{ARRS99}. 
This front displays an intriguing staircase-like fine-structure in the density 
\cite{HRS04} which has been also observed in the fronts of various
quantum spin chains and in equivalent fermionic systems
\cite{GKSS05,PK07,LM10,BG12}. For free 
fermions, the fine structure appears to be related to the curvature of
the single-particle dispersion around the Fermi points and thus cannot
be accounted for in the usual semiclassical picture. Nevertheless, 
the staircase appearance of the front suggests a particle interpretation
\cite{HRS04} and thus the FCS appears as a natural candidate
for exploring the details of quantum fronts.

Our aim here is to probe the quantum noise in the front region by 
developing the FCS in a frame comoving with the edge of the front.
For the free fermion case, our main result is the discovery of
a connection between the FCS
at the front edge and the distribution functions of the
largest eigenvalues of the Gaussian unitary random matrix ensemble.
This result allows us to characterize the front in terms of 
particles whose statistics, and in particular, their order
statistics can be derived exactly. 

\begin{figure}[htb]
\center
\includegraphics[width=\columnwidth]{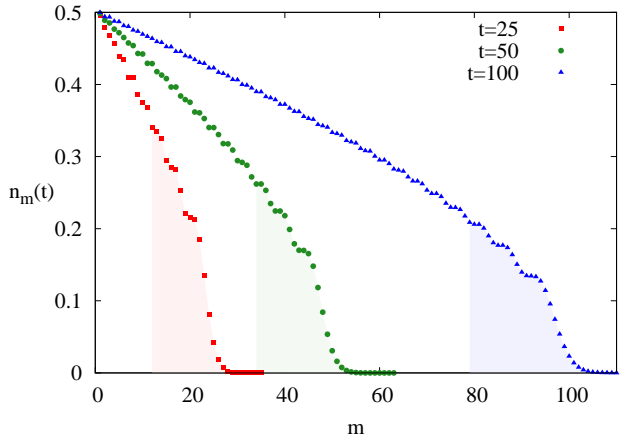}
\caption{Density profiles $n_m(t)$ at increasing times. The edge of the front with the
emerging staircase structure is shown by the shaded region. The profile for $m\le 0$
is given by $n_m(t) = 1-n_{1-m}(t)$ and is not shown.}
\label{fig:dprof}
\end{figure}

The system under consideration is an infinite chain of free spinless fermions
described by the Hamiltonian
\eq{
\hat H =- \frac 1 2 \sum_{m=-\infty}^{\infty}
(c_m^{\dagger} c_{m+1} + c_{m+1}^{\dagger} c_{m})
}
where $c^{\dagger}_m$ is a fermionic creation operator at site $m$. Initially,
all the sites with $m \le 0$ are filled while those with $m >0$ are empty and thus the
correlations at time $t=0$ are given by
\eq{
\langle c^{\dagger}_m c_n\rangle = 
\left\{
\begin{array}{ll}
\delta_{mn} & m,n \le 0 \\
0 & \mbox{else}
\end{array}\right. .
}
%
%= \sum_q \omega_q c_q^{\dagger}c_q
Since $\hat H$ is diagonalized by a Fourier transform with a single-particle spectrum
$\omega_q = -\cos q$, the time evolution of the Fermi operators $c_m(t)$ can be obtained
in terms of Bessel functions $J_m(t)$ as
\eq{
c_m(t) = \sum_{j=-\infty}^{\infty} i^{j-m} J_{j-m}(t) \, c_j \, .
\label{eq:cqt}}
The step function in the density $n_m(t)$ spreads out ballistically and for $m>0$
is given by \cite{ARRS99}
\eq{
n_m(t) = \langle c^{\dagger}_m(t) c_m(t)\rangle = \frac{1}{2}(1-J_0^2(t)) - \sum_{k=1}^{m-1} J_k^2(t)
\label{eq:nmt}}
with snapshots of the density profile shown on Fig. \ref{fig:dprof}. The curves for different times
collapse in the bulk by choosing the scaling variable $m/t$, and the emerging profile can
be understood by semiclassical arguments \cite{AKR08}.
However, there is a nontrivial staircase structure emerging around the edge of the front,
shown by the shaded areas in Fig. \ref{fig:dprof}. The size of the edge region was found to scale with
$t^{1/3}$ and it was argued that each step contains one particle \cite{HRS04}.
Since the origin of this behaviour cannot be captured by a semiclassical treatment,
a more detailed understanding will be obtained by looking at the FCS which includes
all the higher order correlations of the particle number.

%%%%%%%%%%%%%%%%%%%%%%%%%%%%%%%%%%%%%%%%%%%%%%%%%%%%%%%%%%%
%
%
%%%%%%%%%%%%%%%%%%%%%%%%%%%%%%%%%%%%%%%%%%%%%%%%%%%%%%%%%%%

The FCS is defined through the generating function
\eq{
\chi(\lambda,t) = \langle \exp ( i \lambda \hat N_A(t) )\rangle
}
where $\hat N_A(t) = \sum_{m \in A} c^{\dagger}_m (t) c_m(t)$ is the particle number operator
in subsystem $A$ at time $t$. The subsystem will be chosen to include
only the edge of the front and thus $A$ itself is time dependent.
%(the right hand side of $A$ extends to infinity).
The generating function can be written as a determinant \cite{Klich03,Schoenhammer07}
\eq{
\chi(\lambda,t) =  \det \left[ {\bf 1}+(\ee^{i\lambda}-1) {\bf C} (t) \right]
\label{eq:chilt}
}
where ${\bf C}(t)$ is the \emph{reduced} correlation matrix at time $t$ with
matrix elements $C_{mn}(t)=\langle c^{\dagger}_m(t) c_n(t)\rangle$ restricted to the subsystem,
$m,n \in A$, while ${\bf 1}$ is the identity matrix on the same interval.
For the step initial condition one obtains the matrix elements as \cite{ARRS99}
\eq{
C_{mn}(t) = \frac{i^{n-m} \, t}{2(m-n)} \left[J_{m-1}(t)J_{n}(t) - J_{m}(t) J_{n-1}(t) \right]
\label{eq:cmn}}
%
%The phase factors $i^{n-m}$ can be eliminated by 
For convenience, we define $\tilde C_{mn}(t)$ by dropping the phase factor $i^{n-m}$ in Eq. (\ref{eq:cmn})
which corresponds to a simple unitary transformation of the matrix ${\bf C}(t)$ and thus leaves
$\chi(\lambda,t)$ invariant.
Note, that $\tilde {\bf C}(t)$ has exactly the same form as the ground-state correlation matrix of a free fermion chain
with a gradient chemical potential \cite{EIP09}. Thus our results can be directly applied to the static
interface problem as well.

In order to explore the edge region of the front which scales as $t^{1/3}$,  we introduce scaling variables
$x$ and $y$ through $m=t+2^{-1/3}t^{1/3}x$ and $n = t+2^{-1/3}t^{1/3}y$.
We also use the identity
$J_{m-1}(t) = \dot{J}_m(t) + \frac{m}{t}J_m(t)$
to rewrite $\tilde C_{mn}(t)$ in terms of Bessel functions and their time derivatives as
\eq{
%\begin{align}
%\tilde C_{mn}(t) & = 
\frac{t}{2(m-n)} \left[\dot J_m(t) J_n(t) - J_m(t)\dot J_n(t) \right] + \frac{1}{2} J_m(t) J_n(t)
\label{eq:cmn2}
%\end{align}
}
%
%We are interested in the edge region of the front, the size of which scales as $t^{1/3}$.
In the asymptotic regime $t \to \infty$,
one has the following forms \cite{AbrSteg} for the Bessel function and its time derivative
\eq{
\begin{split}
&J_m(t) \approx 2^{1/3} t^{-1/3} \Ai (x) \\
&\dot J_m(t) \approx -2^{2/3}t^{-2/3} \Ai' (x)
\end{split}
}
and the other pair of equations is obtained by interchanging $m \to n$ and $x \to y$.
Substituting into Eq. (\ref{eq:cmn2}) one arrives at
\eq{
\tilde C_{mn}(t) \approx 2^{1/3}t^{-1/3} K(x,y) +
2^{-1/3}t^{-2/3} \Ai (x) \Ai(y)
%+ \mathcal{O}(t^{-1})
\label{eq:cmn3}}
where
\eq{
K(x,y)=\frac{\Ai(x)\Ai'(y)-\Ai'(x)\Ai(y)}{x-y} \, .
\label{eq:kxy}}
%
%The first term in Eq. (\ref{eq:cmn3}) can be written as $K(x,y) \dd y$, thus the factor $2^{1/3}t^{-1/3}$
%accounts for the change of variables while the other term vanishes in the scaling limit.
The  factor $2^{1/3}t^{-1/3}$ in the first term of Eq. (\ref{eq:cmn3})
accounts for the change of variables while the other term vanishes in the scaling limit.
Therefore, the reduced correlation matrix is turned into an integral operator and the generating function
is given by the Fredholm determinant
\eq{
\chi (\lambda,s) = \det \left[ 1+(\ee^{i\lambda}-1) K \right]
\label{eq:chils}}
where $s$ denotes the scaled left endpoint of the domain $A = \left( s, \infty \right) $
over which the kernel $K(x,y)$ is defined.

The Airy kernel (\ref{eq:kxy}) is well known in the theory of random matrices from the
Gaussian unitary ensemble (GUE). Namely, it appears in the level spacing
distribution at the edge of the spectrum \cite{TW94,Mehta}. The bulk eigenvalue density of $N \times N$
GUE matrices is described by the so-called Wigner semicircle with the endpoints at $\pm \sqrt{2N}$.
However, in order to capture the fine structure of the edge of the spectrum one has to magnify it
by choosing the scaling variable $\sqrt{2N} + 2^{-1/2} N^{-1/6} s$.
Then the probability that exactly $n$ eigenvalues lie in the interval $\left(s,\infty\right)$ is
given by the expression \cite{TW94,Mehta}
\eq{
E(n,s) = \left. \frac{(-1)^n}{n!} \frac{\dd^n}{\dd z^n} \det (1-z K) \right|_{z=1} .
\label{eq:ens}}

%one is naturally led to an analogy between the edge eigenvalues of a GUE random matrix and the
%ermionic particles buliding up the edge of the front.
%In fact, the analogy can be made rigorous.
We can easily prove now that the eigenvalue statistics (\ref{eq:ens}) at the edge of the GUE spectrum
is \emph{identical} to that of the quantum front with $n$ being the number of particles.
Indeed, the generating function of (\ref{eq:ens}) is obtained by a Fourier transform and can
be written as
\eq{
\chi(\lambda,s) = \left. \exp \left( -\ee^{i\lambda} \frac{\dd}{\dd z} \right)
\det (1-z K) \right|_{z=1} .
\label{eq:chils2}}
Here, one can notice the operator of the translation which
acts as $\exp(a \frac{\dd}{\dd z}) f(z) = f(z+a)$ on any analytical function.
Translating the argument and setting $z=1$ afterwards, one immediately recovers
the previous result (\ref{eq:chils}). We have thus mapped the FCS of the quantum
front to the well studied problem of GUE edge eigenvalue statistics.
One should point out, however, that while the edge region of the front widens with
time, the edge of the random matrix spectrum shrinks with the matrix size and
thus the correspondence is valid only after proper rescaling.

As an application of the mapping, we revisit the question 
of interpretation of the edge density profile in terms 
of single particles \cite{HRS04}, and we also obtain the 
order statistics of the particles. In the random matrix language,
the probability \emph{density} of the $n$th largest
eigenvalue for GUE random matrices is given by \cite{TW94}
\eq{
F(n,x)=\sum_{k=0}^{n-1}\frac{\dd E(k,x)}{\dd x} \quad .
\label{eq:fns}
}
In addition, one has the following sum rules \cite{Mehta}
\eq{
\sum_{k=0}^{\infty} E(k,x) =1, \quad
\sum_{k=0}^{\infty} k E(k,x) = \Tr K
\label{eq:sumrule}
}
with obvious probabilistic interpretations. 
Using the above sum rules, one arrives at the scaled density of eigenvalues
as a sum of single eigenvalue densities
\eq{
\sum_{n=1}^{\infty} F(n,x) = -\sum_{k=0}^{\infty} k \frac{\dd E(k,x)}{\dd x} = \rho(x)
\label{eq:sumf}
}
where 
\eq{
\rho(x)=K(x,x) = (\Ai'(x))^2 - x \Ai^2(x)
\label{eq:kxx}}
is obtained from Eq. (\ref{eq:kxy}) by taking the limit $x \to y$.

%
%%%%%%%%%%%%%%%%%%%%%%%%%%%%%%%%%%%%%%%%%%%%%%%%%%%%%%%%%%%
%
\begin{figure}[htb]
\center
\includegraphics[width=\columnwidth]{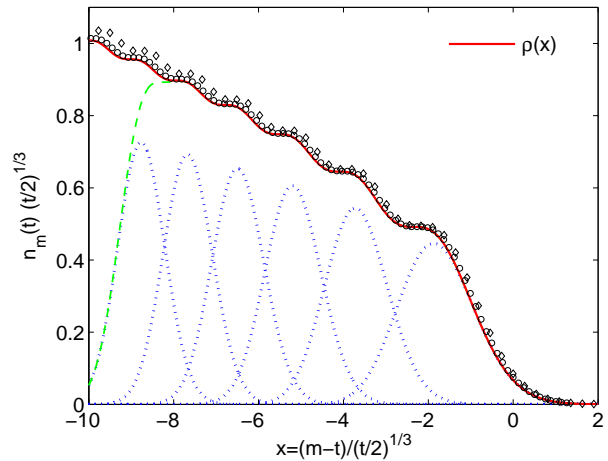}
\caption{Rescaled density near the edge of the front for $t=100$ (diamonds) and $t=1000$ (circles).
The solid (red) line shows the scaling function $\rho(x)$ while the distribution functions
of the $n$th particle for $n=1,\dots,6$ are shown by the dotted (blue) lines.
The dashed (green) line is the sum of the $F(n,x)$.
}
\label{fig:nthld}
\end{figure}
%
%%%%%%%%%%%%%%%%%%%%%%%%%%%%%%%%%%%%%%%%%%%%%%%%%%%%%%%%%%%

The results (\ref{eq:fns}) and (\ref{eq:sumf}) have clear interpretations in the particle picture.
Namely, the probability distribution of the  $n$th eigenvalue $F(n,x)$ is the probability 
distribution of the scaled position of the $n$th particle. Thus the functions $F(n,x)$
provide us with the {\it order statistics} of the rightmost particles in the front with $F(1,x)$
being the Tracy-Widom distribution \cite{TW94}. 
Furthermore, the sum of the probability distributions $F(n,x)$ gives us the scaled density $\rho(x)$
of particles in the front. Fig. \ref{fig:nthld} shows the distributions of the first six particles obtained
by a powerful numerical method for the evaluation of Fredholm determinants \cite{Bornemann10}.
The sum of them, shown by the dashed line, deviates from $\rho(x)$ only at $x \approx -8$.

It is remarkable that the description of the front we found 
is consistent with a classical particle picture. One should also remember, 
however, that the position of the particles spreads out as $t^{1/3}$ in the original (unscaled)
variables.  This subdiffusive spreading has a quantum origin, namely it follows from 
the cubic nonlinearity of the dispersion around the Fermi points.

Additional information on the front region is contained in the cumulants $\kappa_n$ of the FCS.
They can be obtained from logarithmic derivatives
$\kappa_n = \left. (-i \partial_\lambda)^n \ln \chi (\lambda,s) \right|_{\lambda=0}$
of the generating function (\ref{eq:chils}).
Using properties of Fredholm determinants, the cumulant generating function reads
\eq{
\ln \chi (\lambda,s) = \sum_{k=1}^{\infty} (-1)^{k+1} \frac{\Tr K^k}{k} (\ee^{i\lambda}-1)^k
\label{eq:cgf}}
where the trace is defined as
\eq{
\Tr K^k = \int_{s}^{\infty} \dd x_1 \dots \dd x_k K(x_1,x_2) \dots K(x_k,x_1)
\label{eq:trkk}}
The first two cumulants, corresponding to the total number and to
the fluctuations of the particle number, have simple forms $\kappa_1 = \Tr K$
and $\kappa_2 = \Tr K(1-K)$, respectively. In general, carrying
out the traces in $\kappa_n$ is difficult and would require knowledge of
the eigenvalues of the kernel. Having the spectrum would also give access to more
complicated quantities of interest such as the entanglement entropy \cite{KL09,Song11,CMV12}
\eq{
S = - \Tr \left[ K \ln K + (1-K) \ln (1-K) \right] .
\label{eq:ent}}
Even though there exists a differential operator commuting with $K$ \cite{TW94},
the solution of its eigenproblem is not known and the analytic calculation of
$\kappa_2$ and $S$ remains a challenge.

\begin{figure}[htb]
\center
\includegraphics[width=\columnwidth]{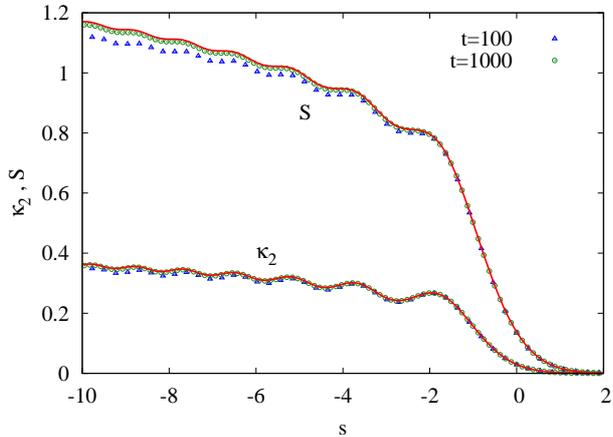}
\caption{Entanglement entropy S (upper symbols) and particle-number fluctuations $\kappa_2$ (lower symbols)
in the edge region for different times as functions of the scaled coordinate $s$.
The solid (red) lines are the $t \to \infty$ scaling functions.
}
\label{fig:entfluct}
\end{figure}

Numerically, we can calculate $\kappa_2$ and $S$ from the matrix representation
of the FCS given in (\ref{eq:chilt}), using matrices of sufficiently large size.
The results are shown in Fig. \ref{fig:entfluct} for different times, plotted against the 
scaling variable $s$. One can see that the convergence to the $t\to\infty$ 
limit is fast. Indeed, for $t=1000$, we have a nearly perfect collapse onto the
scaling functions $\kappa_2=\Tr K(1-K)$ and $S$ given by Eq. (\ref{eq:ent})
and evaluated by using the methods described in \cite{Bornemann10}. This gives us a 
further check on the scaling function of the FCS (\ref{eq:chils}). One should also note 
that both $\kappa_2$ and $S$ inherits signatures of the discrete particle
picture developed above.

%%%%%%%%%%%%%%%%%%%%%%%%%%%%%%%%%%%%%%%%%%%%%%%%%%%%%%%%%%%
%
%
%%%%%%%%%%%%%%%%%%%%%%%%%%%%%%%%%%%%%%%%%%%%%%%%%%%%%%%%%%%

We have found a direct correspondence between counting statistics of free fermions
at the edge of an evolving quantum front and that of random matrix eigenvalues in GUE edge spectra.
A similar correspondence between the equilibrium FCS of fermions in a \emph{line segment}
and the statistics of \emph{bulk} eigenvalues in the GUE spectrum can also be recognized.
Indeed, in a proper continuum limit, the FCS can again be written as the Fredholm determinant (\ref{eq:chils})
with the Airy kernel $K(x,y)$ replaced by the \emph{sine kernel} \cite{AIQ11,IAC11} which is associated with
the bulk spectra of GUE matrices.
For fermions on a finite chain, the connection to random matrices was already pointed out in \cite{KM05}
where the Toeplitz determinant arising in the study of entanglement entropy was related
to the eigenvalue statistics of the circular unitary ensemble (CUE).
Note, however, that the CUE and the GUE statistics become identical in the bulk scaling limit \cite{Mehta}.

Interestingly, the equilibrium statistics also appears in the context of our non-equilibrium problem
if one considers a finite segment in the middle of the chain ($m,n \ll t$) in the limit $t \to \infty$.
Then the matrix elements $C_{mn}(t)$ become independent of $t$ and unitarily equivalent to the equilibrium
correlations \cite{ARRS99}, as can be seen from an asymptotic expansion of the Bessel functions in (\ref{eq:cmn}).
As discussed in the previous paragraph, this implies bulk GUE statistics for the FCS.

It would be important to check the universality of the edge behaviour of quantum fronts. The simplest
generalization would be to start from an initial state, where the left (right) hand side of the chain is
not completely filled (empty) \cite{HRS04} or the initial density profile is a smooth function \cite{BG12}.
In such cases the staircase structure has been found to be essentially unchanged.
Moreover, a very similar structure of the magnetization front was observed in the transverse Ising model
starting from a domain wall initial state \cite{PK05}. One might expect that the FCS remains unchanged in
the above cases.

Another important question is the role of interactions. Does the edge structure survive
if one remains in the integrable Luttinger liquid regime? How robust is it against
integrability breaking terms? Recently, a numerical technique was proposed
\cite{Zauner12} which might be suitable to attack these questions.

Finally, we mention an intriguing connection to the asymmetric exclusion process
\cite{Johansson00,PS00,Sasamoto07}. There, starting from a step initial condition,
the distribution of  particle positions can also be calculated \cite{TW09}.
One observes a $t^{1/2}$ scaling for the width of the distributions at the edge of the front,
while the $t^{1/3}$ scaling and the Tracy-Widom distribution emerge towards the bulk where
the exclusion interaction between the particles becomes more important.
Although there is no one-to-one correspondence to our result, nevertheless one may ask
whether the quantum effects due to the curvature in the spectrum could be described in terms
of some generalized semiclassical picture, e.g. by introducing effective interactions between
particles of different velocities.

This work was supported by the ERC grant QUERG and by the Hungarian Academy of Sciences
through OTKA Grants No. K 68109, NK 100857. We are indebted to Folkmar Bornemann
for his help in using his Matlab toolbox for the numerical evaluation of distributions in
random matrix theory \cite{Bornemann10}.

\bibliographystyle{apsrev4-1.bst}

\bibliography{frontedge}

\end{document}